\documentclass[aoas,preprint]{imsart}
\RequirePackage[OT1]{fontenc}
\RequirePackage{amsthm,amsmath}
\RequirePackage{natbib}
\RequirePackage[colorlinks,citecolor=blue,urlcolor=blue]{hyperref}

\startlocaldefs
\numberwithin{equation}{section}
\theoremstyle{plain}

\theoremstyle{remark}
\newtheorem{note}{Note}
\endlocaldefs

\usepackage{graphicx}
\usepackage[margin=1in]{geometry}                
\allowdisplaybreaks

\begin{document}

\begin{frontmatter}
\title{A Note on the Specific Source Identification Problem in Forensic Science in the Presence of Uncertainty about the Background Population}
\runtitle{Specific Source Identification}

\begin{aug}
\author{\fnms{Danica M.} \snm{Ommen}\thanksref{m1}\ead[label=e1]{danica.ommen@sdstate.edu}},
\author{\fnms{Christopher P.} \snm{Saunders}\thanksref{m1}\ead[label=e2]{christopher.saunders@sdstate.edu}}
\and
\author{\fnms{Cedric} \snm{Neumann}\thanksref{m1}
\ead[label=e3]{cedric.neumann@sdstate.edu}
\ead[label=u1,url]{http://www.sdstate.edu/mathstat/}}

\runauthor{D. Ommen et al.}

\affiliation{South Dakota State University\thanksmark{m1}}

\address{
Department of Mathematics and Statistics \\
South Dakota State University \\
Brookings, SD 57007\\
\printead{e1}\\
\phantom{E-mail:\ }\printead*{e2} \\
\phantom{E-mail:\ }\printead*{e3} \\
\printead{u1}}
\end{aug}

\begin{abstract}
A goal in the forensic interpretation of scientific evidence is to make an inference about the source of a trace of unknown origin.  The evidence is composed of the following three elements: (a) the trace of unknown origin, (b) a sample from a specific source, and (c) a collection of samples from the alternative source population.  The inference process usually considers two propositions. The first proposition is usually referred to as the prosecution hypothesis and states that a given specific source is the actual source of the trace of unknown origin. The second, usually referred to as the defense hypothesis, states that the actual source of the trace of unknown origin is another source from a relevant alternative source population.  One approach is to calculate a Bayes Factor for deciding between the two competing hypotheses.  This approach commonly assumes that the alternative source population is completely known or uses point estimates for its parameters.  Contrary to this common approach, we propose a development that incorporates the uncertainty on the alternative source population parameters in a reasonable and coherent manner into the Bayes Factor.  We will illustrate the resulting effects on the calculation of several Bayes Factors for different situations with a well-studied collection of samples of glass fragments. 
\end{abstract}



\end{frontmatter}

\section{The Identification of Source Problem}
The goal of a forensic scientist is help decide between two competing forensic hypotheses, one presented by the prosecution, denoted $H_p$, and one by the defense, denoted $H_d$. Sampling models corresponding to each of the forensic hypotheses are denoted $M_p$ and $M_d$, respectively.  The statement of the forensic hypotheses and the sampling models depends upon the source identification question being asked to the scientist.  The specific source identification question considers whether the trace originates from a fixed specific source.  For the specific source identification problem, the corresponding forensic hypotheses and sampling model statements are given below\footnote{The specific source identification problem can be contrasted with the common source identification problem which seeks to answer the question of whether or not two traces of unknown origin share a common source.  In the common source case, the source is not fixed.}. 

\begin{description}
\item[\textit{Forensic Hypotheses}]
\footnote{The forensic hypotheses for the common source problem are given by
\begin{description}
\item $H_p$: The two traces originated from the same source.
\item $H_d$: The two traces originated from two different sources.
\end{description}}

\begin{description}
\item $H_p$: The trace originated from the specific source.
\item $H_d$: The trace did not originate from the specific source, but from another source in the relevant alternative source population.
\end{description}

\pagebreak

\item [\textit{Sampling Models}]
\footnote{The sampling models for the common source problem are given by
\begin{description}
\item $M_p$: The two traces were generated by the same randomly selected source in the relevant alternative source population.
\item $M_d$: The two traces were generated by two different randomly selected sources in the relevant alternative source population.
\end{description}}

\begin{description}
\item $M_p$: The trace and the control samples were both generated by the specific source.
\item $M_d$: The trace was not generated by the specific source, but generated by some other randomly selected source in the relevant alternative source population.
\end{description}
\end{description}

Two types of information exist in a case to support either the defense hypothesis or the prosecution hypothesis, quantifiable and conceptual [\cite{AitkenTaroni}]. The conceptual information includes relevant background and circumstantial information, and is denoted $I$.  All the pieces of quantifiable information gathered in relation to a specific trace are collectively called the evidence, denoted $E$. For statistical purposes, $E$ is a set of random elements, whereas $I$ is a set of constraints that determine the form of the sampling models for $E$.

The evidence in the specific source problem is composed of three elements, $E=\{E_u, \ E_s, \ E_a \}$, where $E_u$ denotes a set of observations made on objects from an unknown source, $E_s$ denotes a set of observations made on objects from a fixed specific source, and $E_a$ denotes a set of observations made on objects from a relevant population of possible alternative sources.  For the specific source models, we are assuming that $E_u$, $E_s$, and $E_a$ are three independent samples drawn in the following way:

\begin{enumerate}
\item{$E_{s}$ is a simple random sample from a given specific source determined by $M_p$. Let $\theta_s$ denote the parameters necessary to describe this sampling induced distribution.}
\item{$E_{a}$ is constructed by first taking a simple random sample of sources from a given relevant population of alternative sources; then from each sampled source we have a simple random sample.  Let $\theta_a$ denote the parameters necessary to describe this sampling induced distribution.}
\item{$E_{u}$ is a simple random sample from a single source. It is unknown whether the source of $E_{u}$ is the specific source determined by $M_p$ or if $E_{u}$ typically arises from a randomly selected source in the relevant population of possible alternative sources.
The sampling distribution of $E_{u}$ is characterized by either the parameters $\theta_s$ under $M_p$ or $\theta_a$ under $M_d$.}
\end{enumerate}

Given the background information $I$, it is assumed that $E_s$ contains no further information about $\theta_a$ and conversely, that $E_a$ contains no further information about $\theta_s$.  For the purpose of this paper, we will limit ourselves to the models described above.  When the specific source identification problem deviates from the models described above, for example when $\theta_a$ only contains information on the means of the sources in the alternative population, then the statistical methods will tend to be more complicated than those that follow for this development.  

In most approaches used to compare the prosecution and defense models, point estimates are used for parts, or all, of the parameters describing the distribution of the evidence.  In particular, several authors suggest to calculate Bayes Factors for deciding between the two models [\cite{AitkenStoney}].  However, they assume that $\theta_a$ is known, or they estimate it from the data within a nominal degree of uncertainty.  This approach leads to a Bayes Factor that is intractable and often unfeasible to calculate, especially when the evidence is high-dimensional.  In this paper, we propose a factorization of the Bayes Factor that accounts for the uncertainty on the $\theta_a$ in a reasonable and coherent manner and which can be calculated in practical situations.

\section{Notational Conventions}
We use the following conventions for distinguishing between sampling-induced probability and probability used as a measure of belief:

\begin{enumerate}
\item Latin letters denote sampling induced probability measures; for example, $f(e_s|\theta_s)$ denotes the likelihood of observing the realized value of the sample from the specific source given the actual value of the specific source distribution parameters.
\item Greek letters denote a probability measure that is a measure of belief; for example, $\pi(\theta_s|e_s)$ denotes the posterior density of $\theta_s | e_s$, which describes our belief about the value of $\theta_s$ after observing a sample $E_s|\theta_s$.
\item When combining a belief with a sampling induced probability through Bayes theorem, the result is another belief that is informed or updated by the observed sample. We denote the resulting distribution with a $\pi$. 
\end{enumerate}

In this setting, the stochastic nature of the evidence $E$ is characterized by an unknown but fixed set of parameters $\theta$.  However, $\theta$ is usually of interest only in so far as knowledge of its value facilitates the quantification of support that $E$ provides for either the prosecution hypothesis or the defense hypothesis. In this sense, having to estimate $\theta$ is a nuisance, and hence in the statistical nomenclature these parameters in this situation are known as \textit{`nuisance parameters'}.  The application of Bayesian methods to our problem requires us to specify priors for these nuisance parameters; one summarizing our belief about how the specific source generates evidence, $\pi(\theta_s)$, and another summarizing our prior belief about how the alternative source population stochastically generates evidence, $\pi(\theta_a)$.  

\section{Statistical Methods}

\subsection{Introduction to the Bayes Factor}

Dating back to the 1970's, the specific source identification problem has been approached within the context of subjective Bayesian forensic hypothesis testing (See \cite{AitkenStoney}, \cite{Lindley}, and \cite{Shafer} for more details).  Historically, the specific source identification problem has been limited to applications in which the data is inherently low dimensional, the stochastic nature of the evidence can be characterized by a common parametric family of distributions, and the evidence from the alternative source population is sufficiently precise that it completely characterizes the stochastic nature of the alternative source population. By approaching the specific source identification problem as a Bayesian hypothesis test, the forensic statistician is tasked with providing a summary of the scientific evidence that is logical and coherent for updating a prior belief structure concerning the two competing hypotheses.  The summary is typically known as a Bayes Factor [\cite{Good}]. Traditionally, this summary is used within the context of a Bayes' rule as follows:
\begin{equation}\label{BF}
\underbrace{
\frac
{P\left( H_p | e, I\right)}
{P\left( H_d | e, I\right)}
}_\text{Posterior Odds}  
=
\underbrace{
\frac
{P\left( e |H_p, I \right)}
{P\left( e |H_d, I \right)}
}_{
\begin{smallmatrix}
  \text{Bayes Factor and/or} \\
  \text{Likelihood Ratio}
\end{smallmatrix}
}
\times \ 
\underbrace{
\frac
{P\left( H_p | I \right)}
{P\left( H_d | I \right)}
}_\text{Prior Odds} 
,
\end{equation}

where $e$ is the realization of the evidence, $H_p$ is the prosecution hypothesis, $H_d$ is the defense hypothesis, and $I$ is the relevant background information common to both hypotheses.  The prior odds summarize our relative belief concerning the validity of the prosecution and defense forensic hypotheses before observing the evidence.  

The Bayes Factor then allows us to update our belief following the observation of the evidence and arrive at the posterior odds concerning the relative validity of the two hypotheses. If the Bayes Factor (and the corresponding posterior odds) is sufficiently high, then we support the prosecution hypothesis; on the other hand if it is sufficiently close to zero, we support the defense hypothesis. In effect the Bayes Factor is providing a numerical summary of the answer to both of these questions:

\begin{quote}
\textit{``What is my belief about the likelihood of observing the evidence under the prosecution hypothesis?"}
\end{quote}

versus

\begin{quote}
\textit{``What is my belief about the likelihood of observing the evidence under the defense hypothesis?"}
\end{quote}

It is extremely important note that the use of a Bayes Factor in the context of formal Bayesian model selection, two sets of probability measures are required.  The first is the prior beliefs concerning the relative validity of the two competing models, which has been described as the prior odds in Equation~\ref{BF}.  The second is a set of priors that characterizes the belief about the parameters for the stochastic generation of the evidence under the prosecution and defense models.   Since the main focus of this paper is to study the Bayes Factor for the specific source identification problem under certain conditions, we will only discuss the second set of priors for characterizing the parameters of the sampling models.

\subsection{Known Alternative Source Population Parameters}

As an introduction to derivations associated with the Bayes Factor for the specific source identification problem, we will derive the value of evidence as presented by \cite{Lindley}. In this section we are assuming that we have a well-studied alternative source population with known parameters, $\theta_{a_{_0}}$.  The only unknown parameters that are contributing to the uncertainty about the value of the evidence are the ones associated with the specific source, $\theta_s$.  Let $e=\{e_s, \ e_u, \ e_a \}$ represent the realization of the random element $E$ for a specific case at hand. Let $\theta = \{ \theta_s, \ \theta_{a_{_0}} \}$ and $\pi(\theta)=\pi(\theta_s)$ be a probability distribution used to describe our prior belief about $\theta$ since there is no uncertainty about $\theta_{a_{_0}}$.

Define the value of the evidence as

\begin{equation}\label{Vss}
V_{ss}(e) = \dfrac{\pi(e|H_p, I)}{\pi(e|H_d, I)}.
\end{equation}

Computing the value of evidence in this form involves evaluating the likelihood of the entire set of evidence $e$.  This can be computationally intensive and often unfeasible.  To obtain a computationally tractable form of the value of evidence, \cite{AitkenTaroni} have proposed a factorization of $V_{ss}(e)$ which assumes that $\theta_a$ is known or can be estimated from the data.  The factorization develops as follows:

\begin{align*}
V_{ss}(e) &= \dfrac{\pi(e|M_p)}{\pi(e|M_d)} &\text{see Note~\ref{prfV1} below}
\\ &= \dfrac{\int f(e|\theta, M_p) \pi(\theta|M_p) d\theta}{\int f(e|\theta,M_d) \pi(\theta|M_d) d\theta} &\text{see Note~\ref{prfVmarg} below}
\\ &= \dfrac{\int f(e_u|\theta, M_p) f(e_s|\theta, M_p) f(e_a|\theta, M_p) \pi(\theta) d\theta}{\int f(e_u|\theta, M_d) f(e_s|\theta, M_d) f(e_a|\theta, M_d) \pi(\theta) d\theta} &\text{see Note~\ref{prfV2} below}
\\ &= \dfrac{\int f(e_u|\theta_s) f(e_s|\theta_s) f(e_a|\theta_{a_{_0}}) \pi(\theta_s) d\theta_s}{\int f(e_u|\theta_{a_{_0}}) f(e_s|\theta_s) f(e_a|\theta_{a_{_0}}) \pi(\theta_s) d\theta_s} &\text{see Note~\ref{prfV3.1} below}
\\ &= \dfrac{f(e_a|\theta_{a_{_0}}) \int f(e_u|\theta_s) f(e_s|\theta_s) \pi(\theta_s) d\theta_s}{f(e_u|\theta_{a_{_0}}) f(e_a|\theta_{a_{_0}}) \int f(e_s|\theta_s) \pi(\theta_s) d\theta_s} 
\\ &= \dfrac{1}{f(e_u|\theta_{a_{_0}}) } \times \dfrac{\int f(e_u|\theta_s) f(e_s|\theta_s) \pi(\theta_s) d\theta_s}{\int f(e_s|\theta_s) \pi(\theta_s) d\theta_s} &\text{see Note~\ref{prfV5} below}
\\ &= \dfrac{1}{f(e_u|\theta_{a_{_0}}) } \times \int f(e_u|\theta_s) \dfrac{f(e_s|\theta_s) \pi(\theta_s)}{\int f(e_s|\theta_s) \pi(\theta_s) d\theta_s} d\theta_s
\\ &=  \dfrac{\int f(e_u|\theta_s) \pi(\theta_s|e_s) d\theta_s}{f(e_u|\theta_{a_{_0}})} &\text{see Note~\ref{prfVpostBelief} below}
\\ &= \dfrac{\pi(e_u|e_s, M_p, I)}{f(e_u|\theta_{a_{_0}})} &\text{see Note~\ref{prfVpostPred} below}
\end{align*}

\begin{note}\label{prfV1}
We can drop the conditional notation on $I$ since the background information will be the same for both the prosecution and the defense and the relevant information has been considered in the  models.
\end{note}

\begin{note}\label{prfVmarg}
The definition of the marginal belief of $X$ given some parameter $\phi$ is $\pi(x)=\int f(x|\phi) \pi(\phi) d\phi$.
\end{note}

\begin{note}\label{prfV2}
$f(e|\theta)$ is the likelihood function for observing $e$.  Therefore, $f(e|\theta) = f(e_u|\theta) f(e_s|\theta) f(e_a|\theta)$.  Also, $\pi(\theta)$ does not depend on $M_p$ or $M_d$ so the conditional notation can be dropped.
\end{note}

\begin{note}\label{prfV3.1}
By definition of $\pi(\theta)=\pi(\theta_s)$. The parameters for $E_s$ and $E_a$ are fixed so they will be the same for both $M_p$ and $M_d$.  Under the prosecution model, $M_p$, $E_u$ is characterized by $\theta_s$ since the prosecution believes the specific source is the origin of the trace.  Therefore, $f(e_u|\theta, M_p) = f(e_u|\theta_s)$.  Under the defense model, $M_d$, $E_u$ is completely characterized by $\theta_{a_{_0}}$ since the defense believes the trace came from a source in the alternative source population.  Therefore, $f(e_u|\theta, M_d) = f(e_u|\theta_{a_{_0}})$.
\end{note}

\begin{note}\label{prfV5}
It should be noted that $f(e_a|\theta_{a_{_0}})$ cancels from the value of evidence.  This means that since $\theta_{a_{_0}}$ is known, $e_a$ is irrelevant to the resulting value of the evidence.
\end{note}

\begin{note}\label{prfVpostBelief}
The definition of the posterior belief of $\phi$ given $X$ is $\pi(\phi|x) = \dfrac{f(x|\phi) \pi(\phi)}{f(x)} = \dfrac{f(x|\phi) \pi(\phi)}{\int f(x|\phi) \pi(\phi) d\phi}.$
\end{note}

\begin{note}\label{prfVpostPred}
The definition of posterior predictive belief of $Y$ given $X$ is $\pi(y|x) = \int f(y|\phi)\pi(\phi|x) d\phi.$
\end{note}

By assuming we know (or we are certain that we know) the value $\theta_{a_{_0}}$, the denominator reduces to evaluating the sampling distribution of $e_u$; in effect the denominator does not contain any belief measures when the alternative source population parameters are known.  We will refer to
\begin{equation}\label{VforAknown}
V_{ss}(e|\theta_{a_{_0}}) = \dfrac{\pi(e_u|e_s, M_p, I)}{f(e_u|\theta_{a_{_0}})}
\end{equation}
as the factored form of the specific source value of evidence when $\theta_a$ is known.

\subsection{Unknown Alternative Source Population Parameters}

In this section, we propose a factorization of $V_{ss}$ that does not assume that $\theta_a$ is known, but that is still computationally tractable.  Let $e=\{e_s, \ e_u, \ e_a \}$ represent the realization of the random element $E$ for a specific case at hand.  Let $\theta = \{ \theta_s, \ \theta_a \}$.  Due to the uncertainty in the parameters $\theta$ we will need to characterize our belief about it using the prior.  Let $\pi(\theta)=\pi(\theta_s)\pi(\theta_a)$ be the probability distribution used to describe our prior belief about $\theta$.  Note that we are choosing to restrict ourselves to priors on $\theta_s$ and $\theta_a$ that are \textit{independent of each other}.  Starting from the value of the evidence given by Equation~\ref{Vss}, and using similar methods as the previous case, we can rewrite $V_{ss}$ as follows:
\pagebreak

\begin{align}
V_{ss}(e; \theta_s, \theta_a) &= \dfrac{\pi(e|M_p)}{\pi(e|M_d)}  &\text{see Note~\ref{prfV1} above} \notag
\\ &= \dfrac{\int f(e|\theta, M_p) \pi(\theta|M_p) d\theta}{\int f(e|\theta,M_d) \pi(\theta|M_d) d\theta} &\text{see Note~\ref{prfVmarg} above} \notag
\\ &= \dfrac{\int f(e_u|\theta, M_p) f(e_s|\theta, M_p) f(e_a|\theta, M_p) \pi(\theta) d\theta}{\int f(e_u|\theta, M_d) f(e_s|\theta, M_d) f(e_a|\theta, M_d) \pi(\theta) d\theta} &\text{see Note~\ref{prfV2} above} \notag 
\\ &= \dfrac{\int f(e_u|\theta_s) f(e_s|\theta_s) f(e_a|\theta_a) \pi(\theta) d\theta}{\int f(e_u|\theta_a) f(e_s|\theta_s) f(e_a|\theta_a) \pi(\theta) d\theta} &\text{see Note~\ref{prfV3.2} below} \notag
\\ &= \dfrac{\int f(e_u|\theta_s) f(e_s|\theta_s) \pi(\theta_s) d\theta_s \int f(e_a|\theta_a) \pi(\theta_a) d\theta_a}{\int f(e_s|\theta_s) \pi(\theta_s) d\theta_s \int f(e_a|\theta_a) f(e_u|\theta_a) \pi(\theta_a) d\theta_a} &\text{see Note~\ref{prfV4} below} \notag
\\ &= \dfrac{\int f(e_u|\theta_s) f(e_s|\theta_s) \pi(\theta_s) d\theta_s}{\int f(e_s|\theta_s) \pi(\theta_s) d\theta_s} \times \dfrac{\int f(e_a|\theta_a) \pi(\theta_a) d\theta_a}{\int f(e_a|\theta_a) f(e_u|\theta_a) \pi(\theta_a) d\theta_a} \notag
\\ &= \dfrac{\int f(e_u|\theta_s) f(e_s|\theta_s) \pi(\theta_s) d\theta_s}{\int f(e_s|\theta_s) \pi(\theta_s) d\theta_s} \Bigg/ \dfrac{\int f(e_a|\theta_a) f(e_u|\theta_a) \pi(\theta_a) d\theta_a}{\int f(e_a|\theta_a) \pi(\theta_a) d\theta_a} \notag
\\ &= \dfrac{\int f(e_u|\theta_s) \pi(\theta_s|e_s) d\theta_s}{\int f(e_u|\theta_a) \pi(\theta_a|e_a) d\theta_a} &\text{see Note~\ref{prfVpostBelief} above} \notag
\\ &= \dfrac{\pi(e_u|e_s, M_p)}{\pi(e_u|e_a, M_d)} &\text{see Note~\ref{prfVpostPred} above} \notag
\end{align}

\begin{note}\label{prfV3.2}
The parameters for $E_s$ and $E_a$ are fixed so they will be the same for both $M_p$ and $M_d$.  Under the prosecution model, $M_p$, $E_u$ is characterized by $\theta_s$ since the prosecution believes the specific source is the origin of the trace.  Therefore, $f(e_u|\theta, M_p) = f(e_u|\theta_s)$.  Under the defense model, $M_d$, $E_u$ is characterized by $\theta_{a}$ since the defense believes the trace came from a source in the alternative source population.  Therefore, $f(e_u|\theta, M_d) = f(e_u|\theta_{a})$.
\end{note}

\begin{note}\label{prfV4}
Since $\pi(\theta)=\pi(\theta_s)\pi(\theta_a)$, which means that $\theta_s$ and $\theta_a$ are independent, we can factor the integral apart with respect to $\theta$ into the product of the integrals for $\theta_s$ and $\theta_a$.
\end{note}

We will refer to
\begin{equation}\label{VforAunknown}
V_{ss}(e) = \dfrac{\pi(e_u|e_s, M_p)}{\pi(e_u|e_a, M_d)}\end{equation}
as the factored form of the specific source value of evidence when $\theta_a$ is unknown.

\subsection{Results of the Factored Forms for the Value of Evidence}

When the value of evidence is given by Equation~\ref{Vss}, the computation involves evaluating the likelihood of the entire set of evidence $E$.  This can be very computationally intensive, and often unfeasible.  In the factored forms given by Equation~\ref{VforAknown} and Equation~\ref{VforAunknown}, the computation only involves the posterior predictive belief of the unknown evidence.  Therefore, the factored forms create a computationally feasible method of evaluating the Bayes Factor for the specific source identification problem that can be calculated using Monte Carlo integration techniques [\cite{KassRaf}].  When the specific source value of evidence is factored into the form given by Equation~\ref{VforAknown} (the case when $\theta_{a_{_0}}$ is known or estimated from the data), $e_a$ is irrelevant to the value of the evidence.  This considerably simplifies the computational complexity of the calculation of $V_{ss}$; however, the uncertainty in the estimation of $\theta_a$ is not formally accounted for.  This uncertainty is taken into account when the value of evidence is calculated using the factorization presented in Equation~\ref{VforAunknown} (the case when the parameters for the alternative source population are unknown).  In this case, Monte Carlo integration techniques have to be used for the calculation of the denominator.  This involves an additional layer of computational complexity; however, modern computers can cope with the required number of computations in a reasonable amount of time.

\section{Glass Example}

In order to compare the value of the evidence obtained using both factorizations of $V_{ss}$ we use a collection of samples of glass fragments studied by \cite{AitkenLucy}.  The dataset consists of three classes of windows, with 16, 16, and 30 windows in each class.  There are 5 glass fragments per window.  Following \cite{AitkenLucy}, we consider the logarithm of the measurements for the ratios of elemental compositions on each glass fragment: $\log(Ca/K)$ is represented by the second variable (V2), $\log(Ca/Si)$ is represented by the third variable (V3), and $\log(Ca/Fe)$ is represented by the fourth variable (V4).

\begin{figure}[h!]
\begin{center}
\includegraphics[width=1\textwidth]{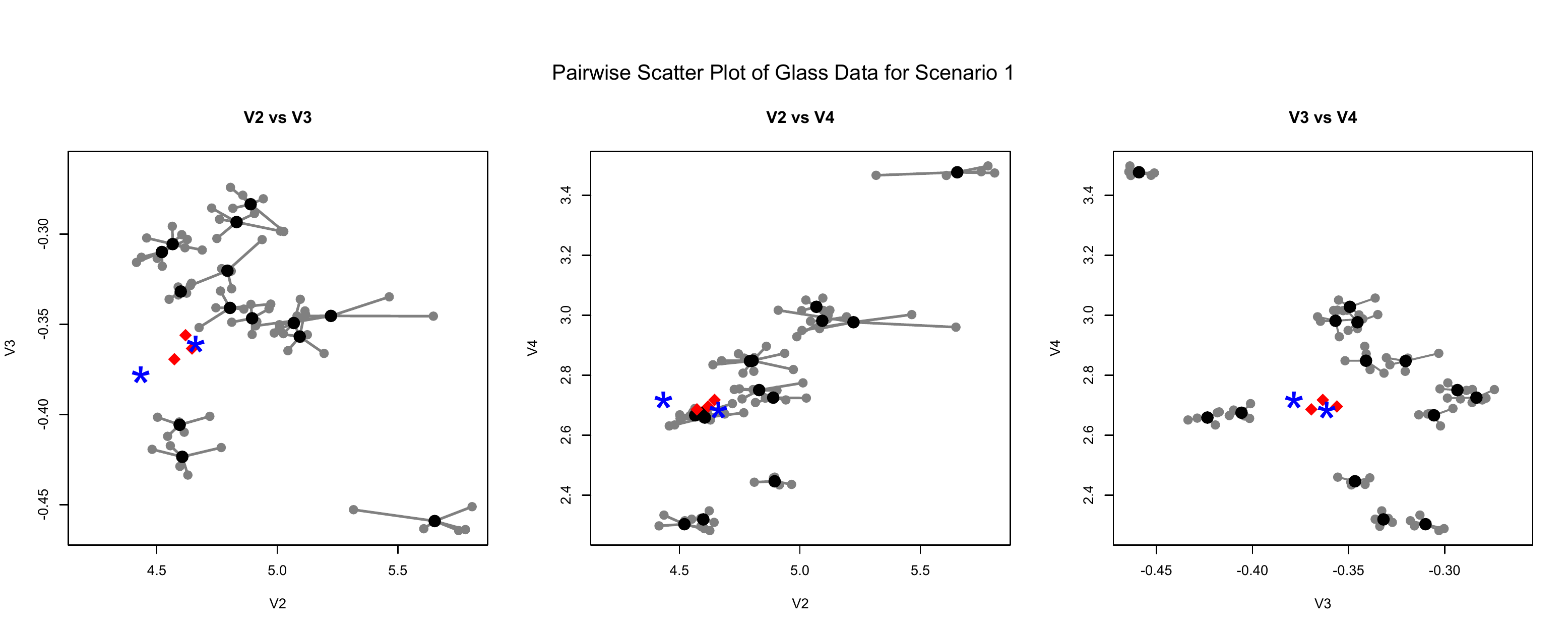}
\includegraphics[width=1\textwidth]{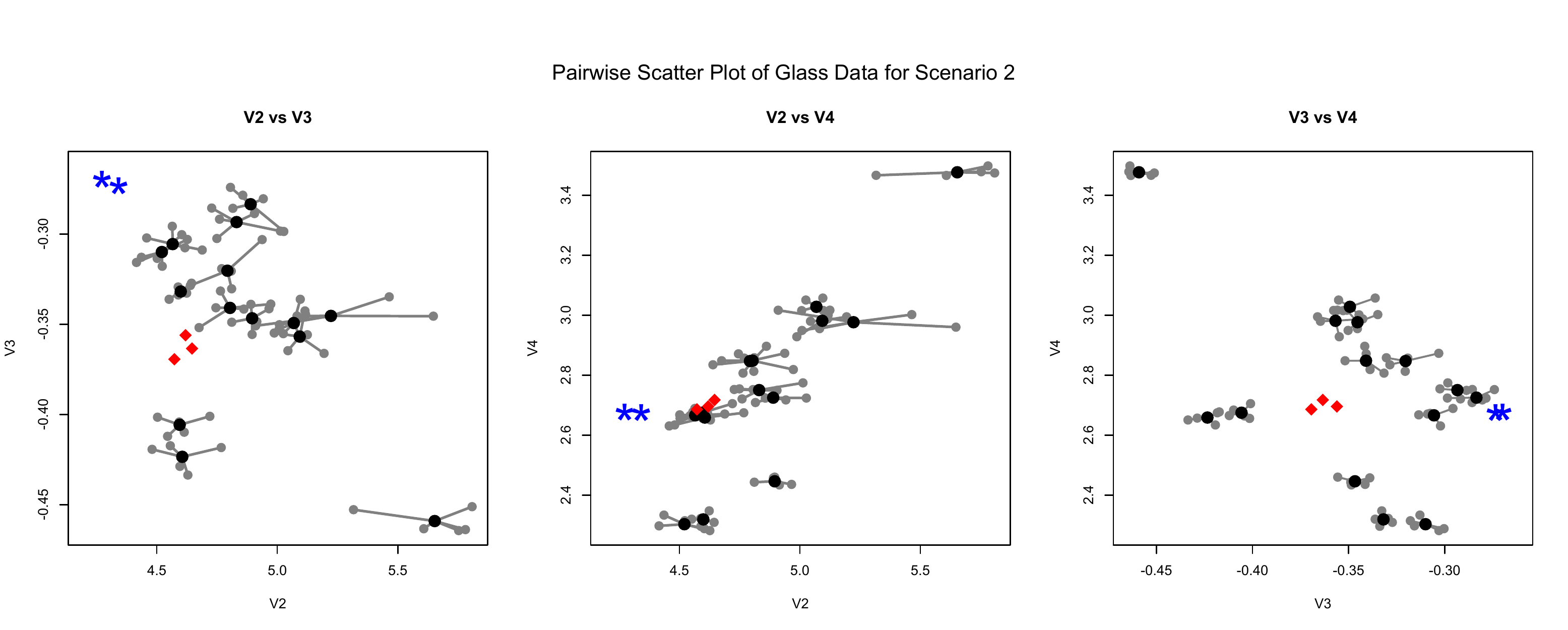}
\caption{\label{GlassPlots}In the pairwise scatterplots of the evidence $E$, the blue asterisks represent $E_u$, the red diamonds represent $E_s$, and the gray dots represent $E_a$.  The large black dots are the mean values for each window and the gray lines show the deviation from that mean for each fragment from the window.} 
\end{center}
\end{figure}

As an illustrative example of computing the value of evidence for the specific source identification problem we focus only on the first class of 16 windows, and we consider two scenarios.  For the first scenario, $e_u$ and $e_s$ will share a fixed source, with the $4^{th}$ window playing the role of the specific source.  The first three fragments from window 4 will serve as $e_s$ and the last two fragments from window 4 will serve as $e_u$.  For the second scenario, $e_u$ and $e_s$ will have different sources.  The first three fragments from window 4 will serve as $e_s$ and two fragments from the $2^{nd}$ window will serve as $e_u$.  In both scenarios, the remaining 70 glass fragments divided among the 14 remaining windows will serve as $e_a$. The pairwise scatter plots of the evidence under each scenario can be seen in Figure~\ref{GlassPlots}.

\vspace{0.05in}
The specific source identification question for this example can stated as
\begin{quote}
\textit{``Did the glass fragments of unknown source come from the fourth window?"}
\end{quote}
The resulting forensic hypotheses are
\begin{description}
\item $H_p$: The glass fragments from the unknown source came from the fourth window.
\item $H_d$: The glass fragments from the unknown source did not come from the fourth window, but from some other window.
\end{description}
The corresponding sampling models are
\begin{description}
\item $M_p$: The glass fragments from the unknown source came from the fourth window. Under this model, $E_u$ can be characterized by the same model which characterizes $E_s$, described in detail below.
\item $M_d$: The glass fragments from the unknown source came from a randomly selected window in the alternative source population.  Under this model, $E_u$ can be characterized by the same model which characterizes $E_a$, described in detail below.
\end{description}

\subsection{Sampling Model for the Evidence from the Specific Source}

First, we will assume that the measurements on the glass fragments composing $e_s$ are an independent and identically distributed, abbreviated i.i.d., sample from a multivariate normal with a mean vector $\mu_s$ and covariance $\Sigma_s$.  Let $y_{sj}$ denote the vector of measurements on the $j^{th}$ fragment from the specific source evidence for $j=1,2, \ldots, m$ (for this example $m=3$), then $y_{sj}$ follows a multivariate normal distribution with a mean vector, $\mu_s$, and a covariance matrix, $\Sigma_s$, which we will denote as $y_{sj} \sim MVN(\mu_s, \Sigma_s)$.  For this example, the specific source population parameters are $\theta_s = \{ \mu_s, \Sigma_s \}$, so we need to specify priors for both $\mu_s$ and $\Sigma_s$.  We will use a relatively non-informative multivariate normal prior on $\mu_s$ with the zero mean vector and a diagonal covariance matrix with diagonal elements equal to $3000$. The prior for $\Sigma_s$ is an Inverse Wishart distribution (denoted $W^{-1}$) centered at a diagonal covariance matrix, $\Phi$ with diagonal elements of $0.01$, $0.00005$, $0.0005$ and three degrees of freedom.  These diagonal elements were chosen based on the approximate precision of the measurements for the evidence.  The full model for $E_s$ with supporting prior beliefs is summarized below.
$$y_{sj} \sim MVN(\mu_s, \Sigma_s)$$
$$\mu_s \sim MVN(0, 3000I)$$
$$\Sigma_s \sim W^{-1}(\Phi, 3)$$

It should be noted that any number of reasonable priors can be chosen for $\mu_s$ and $\Sigma_s$.  The numerator for the values of evidence $\pi(e_u|e_s, M_p)$ under scenario 1 ($Exp \ 1$) and scenario 2 ($Exp \ 2$) are given in Table~\ref{known} and Table~\ref{unknown} below.

\subsection{Sampling Model for the Evidence in the Alternative Source Population}

Next, we assume that the measurements on the glass fragments composing $e_a$ follow a hierarchical multivariate normal model with the assumption that all windows in the alternative source population have a mean $\mu_a$, a within-covariance matrix $\Sigma_w$, and a between source covariance matrix $\Sigma_b$.   Let $y_{ij}$ denote the vector of measurements on the $j^{th}$ fragment, for $j = 1, 2, \ldots, m_i$ (for this example, $m \equiv m_i = 5$ for all $i$) from the $i^{th}$ window, for $i=1, 2, \ldots, n$ (for this example, $n=14$).  The hierarchical multivariate model in this case is a simple random effects model where the between-source effects $a_i$ are i.i.d. multivariate normal random vectors with a mean vector of zero and a covariance matrix  $\Sigma_b$.  The within-source effects $w_{ij}$ are assumed to be i.i.d. multivariate normal vectors with a mean vector of zero and a covariance matrix of $\Sigma_w$.  We will compare the results of both scenarios described above under two different conditions.  First, the alternative source population parameters are assumed to be known.  Secondly, the alternative source population parameters are assumed to be unknown.

When the alternative source population parameters $\theta_a = \{ \mu_a, \Sigma_b, \Sigma_w \}$ are assumed to be known, we use the estimates for the parameters as suggested in \cite{AitkenLucy}.  The estimates $\hat{\theta}_a = \{ \hat{\mu}_a, \hat{\Sigma}_b, \hat{\Sigma}_w \}$ of the known parameters are summarized below.

$$\hat{\mu}_a = \dfrac{1}{mn} \sum_{i=1}^{n} \sum_{j=1}^m y_{ij}$$
$$\bar{y}_i = \dfrac{1}{m} \sum_{j=1}^m y_{ij}$$
$$\hat{\Sigma}_w = \dfrac{1}{n(m-1)} \sum_{i=1}^n \sum_{j=1}^m (y_{ij}-\bar{y}_i)(y_{ij}-\bar{y}_i)^T$$
$$\hat{\Sigma}_b = \left[ \dfrac{1}{n-1} \sum_{i=1}^n (\bar{y}_i-\hat{\mu}_a)(\bar{y}_i-\hat{\mu}_a)^T \right] - \left[ \dfrac{1}{m}\hat{\Sigma}_w \right]$$

The denominator for the values of evidence $f(e_u|\hat{\theta}_{a_{_0}})$ under scenario 1 ($Exp \ 1$) and scenario 2 ($Exp \ 2$) are given in Table~\ref{known} below.

\begin{table}[h!]
\centering
\caption{\label{known}Alternative Source Population Parameters Known}                         
\begin{tabular}{| l | c | c |}
\hline
& $Exp \ 1$ & $Exp \ 2$ \\
\hline
$\pi(e_u|e_s, M_p)$ & $119740.3$ & $2.316277$ \\
$f(e_u|\hat{\theta}_{a_{_0}})$ & $582.6974$ & $144.1683$ \\
$V_{ss}(e|\hat{\theta}_{a_{_0}})$ & $205.4931$ & $0.01606648$ \\
\hline
\end{tabular}
\end{table} 

When the alternative source population parameters $\theta_a = \{ \mu_a, \Sigma_b, \Sigma_w \}$ are unknown, our prior for $\Sigma_w$ is the same that is used for $\Sigma_s$, and the prior for $\mu_a$ is the same as that used for $\mu_s$. We use an Inverse Wishart prior for $\Sigma_b$ centered at the identity covariance matrix with three degrees of freedom.  The full model for $E_a$ with supporting prior beliefs is summarized below.
$$\text{For } i=1, 2, \ldots, 14 \text{ and }j = 1, 2, 3, 4, 5 :$$

\begin{center}                      
\begin{tabular}{ccc}
$y_{ij} = \mu_a + a_i + w_{ij}$ && $\mu_a \sim MVN(0, 3000I)$ \\
$a_i \overset{iid}{\sim} MVN(0, \Sigma_b)$ && $\Sigma_b \sim W^{-1}(I, 3)$ \\
$w_{ij} \overset{iid}{\sim} MVN(0, \Sigma_w)$ && $\Sigma_w \sim W^{-1}(\Phi, 3)$
\end{tabular}
\end{center}

The denominator for the values of evidence $\pi(e_u|e_s, M_p)$ under scenario 1 ($Exp \ 1$) and scenario 2 ($Exp \ 2$) are given in Table~\ref{unknown} below.

\begin{table}[h!]
\centering
\caption{\label{unknown}Alternative Source Population Parameters Unknown}                         
\begin{tabular}{| l | c | c |}
\hline
& $Exp \ 1$ & $Exp \ 2$ \\
\hline
$\pi(e_u|e_s, M_p)$ & $119740.3$ & $2.316277$ \\
$\pi(e_u|e_a, M_d)$ & $30.17140$ & $209.5902$ \\
$V_{ss}(e)$ & $3968.669$ & $0.01105146$ \\
\hline
\end{tabular}
\end{table} 

The computations were performed on a 2012 MacBook Pro with an OS X 10.8.5 operating system, 2.3 GHz Intel Core i7 processor, and 16 GB, 1600 MHz memory using R version 3.0.2.  All posterior predictive distributions for the parameters are estimated using the ``MCMCglmm" package in R [\cite{mcmcglmm}].  Using the parameter values sampled from these posterior predictive distributions, we estimated the posterior predictive beliefs of $e_u$ for the values of evidence using a standard Monte Carlo average integration technique.  The Monte Carlo estimates were based on a sample size of $29,000$ ($30,000$ iterations with a burn-in of $1000$).  We also made use of the matrix form of the multivariate simple random effects models as presented by \cite{Miller} (See the Appendix for details).

\section{Discussion}

In the illustrative example described in Section 4, the behavior of $V_{ss}$ is consistent between the calculations when the alternative source population parameters are assumed to be known (results for this experiment can be found in Table~\ref{known}) and when the alternative source population parameters are unknown (results for this experiment can be found in Table~\ref{unknown}).  The value of evidence for the first scenario suggests in both cases that the evidence is more likely to have been generated according to the prosecution model than by the defense model (since the values of evidence are significantly greater than one); the value of the evidence for the second scenario suggests in both cases that the evidence is more likely to have arisen under the defense model.  These results were expected by the design of the experiment.  However, by contrasting the likelihood of the evidence under the defense model in Table~\ref{known} and Table~\ref{unknown}, we observe that in the first scenario the evidence is more likely when the alternative source population parameters are known than when they are not, while in the second scenario the evidence is more likely when there is uncertainty on the alternative source population parameters. When the number of observations in the alternative source population is small, there is a marked difference between these values which suggests that using the estimates of the parameters is not a reasonable surrogate for calculating the value of the evidence in the presence of uncertainty in the alternative source population parameters.  We expect that the difference between $V_{ss}(e; \hat{\theta}_{a_{_0}})$ and $V_{ss}(e)$ will go to zero as the amount of evidence about the alternative source population becomes large (the rate of convergence is currently being investigated by the authors).  In practice, the use of estimates of the known parameters for the alternative source population may lead to grossly over- or under-estimating the value of the evidence. The direction of the misleading effect will depend upon the rarity of the characteristics of $e_u$ in the alternative source population.  This effect may ultimately mislead the criminal justice system.  

It should be noted that choosing different priors for the model parameters can results in radically different values of evidence.  Additionally, when the alternative source population parameters are unknown there is less freedom in choosing the priors than when the parameters are known.  In order for the factorization of the value of evidence to hold when there is uncertainty in the alternative source population parameters, the prior for the specific source parameters, $\pi(\theta_s)$, must be chosen to be independent from the prior for alternative source parameters, $\pi(\theta_a)$.  This precludes the use of the popular ``random man" prior for the specific source parameters in which the specific source is believed to be typical of the alternative source population.

\section{Conclusion}

Computing the value of evidence when it is given by its raw form (Equation~\ref{Vss}) requires evaluating the likelihood of the entire set of evidence.  In most situations, this evaluation is computationally unfeasible.  However, when the value of evidence is given in the factored forms (Equation~\ref{VforAknown} and Equation~\ref{VforAunknown}) it can be approximated using Monte Carlo integration techniques [\cite{KassRaf}]. In the rare setting when there is no uncertainty about the alternative source population parameters (Equation~\ref{VforAknown}), it is not surprising that including $e_a$ in the calculation of the value of evidence has no impact.  However, in traditional forensic settings, there is rarely  sufficient information about the alternative source population parameters to assume that it is possible to estimate them accurately; henceforth, such situations require a novel approach.

In this paper, we develop a logical and coherent method which formally incorporates the uncertainty on the alternative source population parameters into the calculation of the Bayes Factor. This is a major departure from the ad-hoc approaches to include uncertainty about the background population in the value of the evidence that are currently available in the forensic statistics literature. These methods typically entail the construction of a confidence or credible interval for the likelihood ratio represented by $V_{ss}(e; \hat{\theta}_{a_0})$.  By formally incorporating uncertainty about $e_a$ into the value of evidence we can guarantee that the resulting value is statistically rigorous and that the decisions based on it will be admissible in a statistical decision theoretic sense.  To avoid potentially misleading decisions in the court system, the authors suggest replacing the use of ad-hoc methods with the statistically rigorous methods presented here when there is uncertainty in the alternative source population parameters.

\appendix
\section{}

We will summarize a well-known result from Miller concerning the matrix form of the multivariate simple random effects model [\cite{Miller}]. Let $y_{ij}$ denote the $k$-dimensional vector of measurements on the $j^{th}$ component from the $i^{th}$ source for $j=1, 2, \cdots, m_i$ and $i=1, 2, ..., n$.  Then the simple random effects model is given by

\begin{equation*}
y_{ij} = \mu_a + a_i + w_{ij}
\end{equation*}

where $a_i \overset{iid}{\sim} N_k(\mathbf{0}, \Sigma_a)$ and $w_{ij} \overset{iid}{\sim} N_k(\mathbf{0}, \Sigma_w)$ are independent from each other.  It is often useful to think about the random effects model as a combined multivariate normal random vector.  If $y_{ij}$ follows a simple random effects model, then the vector $Y_i=\begin{pmatrix} y_{i1} & y_{i2} & \cdots & y_{i m_i} \end{pmatrix}^T$ has the following distribution:

$$Y_i \overset{iid}{\sim} N_k(\mu_c, \Sigma_c)$$ 

where $\Sigma_c = \begin{pmatrix} \Sigma_a + \Sigma_w & \Sigma_a & \cdots & \Sigma_a
                                         \\ \Sigma_a & \Sigma_a+\Sigma_w & \ddots & \vdots
                                         \\ \vdots & \ddots & \ddots & \Sigma_a
                                         \\ \Sigma_a & \cdots & \Sigma_a & \Sigma_a + \Sigma_w \end{pmatrix}$ and $\mu_c = \begin{pmatrix} \mu_a \\ \mu_a \\ \vdots \\ \mu_a \end{pmatrix}$ for $i=1, 2, \cdots, n$.
                                         
\section*{Acknowledgements}
The research in this article was supported in part by Award No. 2009- DN-BX-K234 awarded by the National Institute of Justice, Office of Justice Programs, US Department of Justice. The opinions, findings, and conclusions or recommendations expressed in this publication are those of the authors and do not necessarily reflect those of the Department of Justice.    




\bibliography{PubBib}

\begin{thebibliography}{9}

\bibitem[\protect\citeauthoryear{Aitken and Lucy}{2004}]{AitkenLucy}
\begin{barticle}[author]
\bauthor{\bsnm{Aitken},~\bfnm{Colin}\binits{C.}} \AND
  \bauthor{\bsnm{Lucy},~\bfnm{David}\binits{D.}}
(\byear{2004}).
\btitle{Evaluation of trace evidence in the form of multivariate data}.
\bjournal{Journal of the Royal Statistical Society. Series C (Applied
  Statistics)}
\bvolume{53}
\bpages{109-122}.
\end{barticle}
\endbibitem

\bibitem[\protect\citeauthoryear{Aitken and Stoney}{1991}]{AitkenStoney}
\begin{bbook}[author]
\bauthor{\bsnm{Aitken},~\bfnm{Colin}\binits{C.}} \AND
  \bauthor{\bsnm{Stoney},~\bfnm{David}\binits{D.}}
(\byear{1991}).
\btitle{The Use of Statistics in Forensic Science}.
\bpublisher{CRC Press}.
\end{bbook}
\endbibitem

\bibitem[\protect\citeauthoryear{Aitken and Taroni}{2004}]{AitkenTaroni}
\begin{bbook}[author]
\bauthor{\bsnm{Aitken},~\bfnm{Colin}\binits{C.}} \AND
  \bauthor{\bsnm{Taroni},~\bfnm{Franco}\binits{F.}}
(\byear{2004}).
\btitle{Statistics and the Evaluation of Evidence for Forensic Scientists},
\bedition{2nd} ed.
\bpublisher{John Wiley and Sons, Ltd.}
\end{bbook}
\endbibitem

\bibitem[\protect\citeauthoryear{Good}{1991}]{Good}
\begin{bbook}[author]
\bauthor{\bsnm{Good},~\bfnm{I.~J.}\binits{I.~J.}}
(\byear{1991}).
\btitle{"Weight of evidence and the Bayesian likelihood ratio" in The Use of
  Statistics in Forensic Science}.
\bpublisher{CRC Press}.
\end{bbook}
\endbibitem

\bibitem[\protect\citeauthoryear{Hadfield}{2010}]{mcmcglmm}
\begin{barticle}[author]
\bauthor{\bsnm{Hadfield},~\bfnm{Jarrod~D.}\binits{J.~D.}}
(\byear{2010}).
\btitle{MCMC methods for Multi--response Generalised Linear Mixed Models: The
  MCMCglmm R Package}.
\bjournal{Journal of Statistical Software}
\bvolume{33}
\bpages{1-22}.
\end{barticle}
\endbibitem

\bibitem[\protect\citeauthoryear{Kass and Raftery}{1995}]{KassRaf}
\begin{barticle}[author]
\bauthor{\bsnm{Kass},~\bfnm{Robert~E.}\binits{R.~E.}} \AND
  \bauthor{\bsnm{Raftery},~\bfnm{Adrian~E.}\binits{A.~E.}}
(\byear{1995}).
\btitle{Bayes Factors}.
\bjournal{Journal of the American Statistical Association}
\bvolume{90}
\bpages{773-795}.
\end{barticle}
\endbibitem

\bibitem[\protect\citeauthoryear{Lindley}{1977}]{Lindley}
\begin{barticle}[author]
\bauthor{\bsnm{Lindley},~\bfnm{Dennis~V.}\binits{D.~V.}}
(\byear{1977}).
\btitle{A problem in forensic science}.
\bjournal{Biometrika}
\bvolume{64}
\bpages{207-213}.
\end{barticle}
\endbibitem

\bibitem[\protect\citeauthoryear{Miller}{1977}]{Miller}
\begin{barticle}[author]
\bauthor{\bsnm{Miller},~\bfnm{John~J.}\binits{J.~J.}}
(\byear{1977}).
\btitle{Asymptotic Properties of Maximum Likelihood Estimates in the Mixed
  Model of the Analysis of Variance}.
\bjournal{Annals of Statistics}
\bvolume{5}
\bpages{746-762}.
\end{barticle}
\endbibitem

\bibitem[\protect\citeauthoryear{Shafer}{1982}]{Shafer}
\begin{barticle}[author]
\bauthor{\bsnm{Shafer},~\bfnm{Glenn}\binits{G.}}
(\byear{1982}).
\btitle{Lindley's Paradox}.
\bjournal{Journal of the American Statistical Association}
\bvolume{77}
\bpages{325-334}.
\end{barticle}
\endbibitem

\end{thebibliography}
\bibliographystyle{imsart-nameyear}

\end{document}